\documentclass{article}
\usepackage{graphicx} % Required for inserting images
\usepackage{amsmath}
\usepackage{amssymb}
\usepackage{mathtools}
\usepackage{epstopdf}
\usepackage{tcolorbox}
\usepackage{shadow}
\usepackage{caption}
\usepackage{wrapfig}
\usepackage{bm}% bold math
\usepackage{bbm}
\usepackage[auth-sc]{authblk}
\newcommand{\be}{\begin{eqnarray}}
\newcommand{\ee}{\end{eqnarray}}

\usepackage{booktabs} % For better-looking tables
\usepackage{siunitx}  % For decimal-aligned numbers
\usepackage{array}    % For better column formatting
\usepackage{float}

\usepackage[super]{natbib}

\title{Functional Information Decomposition:\\A First-Principles Approach to Analyzing Functional Relationships}
\author[1,*]{Clifford Bohm}
\author[1]{Vincent R. Ragusa}
\author[2]{Arend Hintze}
\author[1,5]{Charles Ofria}
\author[1,5]{Emily Dolson}
\author[3,4,5]{Christoph Adami}
\affil[1]{Department of Computer Science and Engineering\protect\\ Michigan State University, East Lansing, MI 48824, USA}
\affil[2]{Department of MicroData Analytics, Dalarna University, Falun, Sweden}
\affil[3]{Department of Microbiology, Genetics, and Immunology, Michigan State University, East Lansing, MI 48824, USA}
\affil[4]{Department of Physics and Astronomy, Michigan State University, East Lansing, MI 48824, USA}
\affil[5]{Program in Evolution, Ecology, and Behavior\protect\\ Michigan State University, East Lansing, MI 48824, USA}
\affil[*]{cliff.bohm@gmail.com}
\begin{document}

\maketitle

%\begin{center}
%\textbf{\textit{Draft Version 0.1}}    
%\end{center}

\begin{abstract}
A central challenge in analyzing multivariate interactions within complex systems is to decompose how multiple inputs jointly determine an output. Existing approaches generally operate on observed probability distributions and can conflate a system’s intrinsic functional logic with statistical artifacts of limited data. As a result, distinct systems can yield identical observations, rendering information decomposition fundamentally underdetermined and obscuring true higher-order interactions.

We introduce Functional Information Decomposition (FID), both a computational and theoretical framework, which defines informational components with respect to a system’s complete input-output mapping, thereby addressing a core cross-scale inference problem: determining how information carried by individual components combines to shape system-level behavior. When the mapping is fully specified, FID provides a unique decomposition into independent and synergistic contributions. Crucially, given only partial observations, FID characterizes the entire space of consistent decompositions by sampling compatible functions, making inferential limits explicit. A complementary geometric perspective clarifies the structural origin of informational components.

We demonstrate FID’s interdisciplinary utility on canonical logical functions, Conway’s Game of Life, and gene-expression–based prediction of cancer drug response, and provide an open-source implementation. By separating functional architecture from observational distribution, FID offers a principled foundation for analyzing multivariate dependence in both fully and partially observed complex systems.
\end{abstract}

\section{Introduction}

Information theory, originating from Shannon’s foundational work on communication systems, has been applied to diverse disciplines such as machine learning, genetics, physics, and neuroscience. A central challenge in applying information theory to complex systems lies in interpreting multivariate dependencies. Shannon defined mutual information as a measure of statistical dependence between two variables~\cite{Shannon1948}. Subsequent extensions of this framework to three or more variables quantify multivariate dependence~\cite{McGill1954, Fano1961}, but at the cost of collapsing distinct higher-order interaction structures into a single scalar quantity. As a result, mutual information alone fails to distinguish how multiple inputs jointly inform an output. This limitation has motivated the development of frameworks for mutual information decomposition, which aim to resolve this ambiguity by partitioning mutual information into interpretable components that reveal the contribution of each input and quantify higher-order interactions. In this work, we introduce Functional Information Decomposition (FID), a framework for information decomposition at the systems level\footnote{In this work, a system is defined as the functional object; a model induced by an observer’s modeling choices—specifically, the selection of variables and their resolution—together with the data collected under those choices. This system reflects, but does not exhaustively represent, the underlying process that generated the data. Whether the model faithfully captures that process is neither assumed nor decidable within the framework. Accordingly, FID assesses only the system as defined: the entity specified by the model and constrained by the data. Formally, a system is the triplet: (1) a set of input variables, (2) an output variable, and (3) a mapping (deterministic or stochastic) between them.}. Where previous methods attempt decomposition from observation alone—treating observed probability distributions as the object of analysis—our approach recognizes a fundamental limitation of this perspective. Observations, while informative, often provide only partial and biased views of a system’s behavior, and are therefore insufficient for systems-level decomposition. In contrast, FID defines informational components with respect to the system's intrinsic logic.

The challenge of decomposing multivariate information has a long history. Early measures such as Watanabe’s total correlation~\cite{Watanabe1960} quantify overall dependency without revealing its distribution among variables, while McGill’s interaction information \cite{McGill1954}---the commonly accepted form for many-way mutual information---captures higher-order effects but can yield negative values whose interpretation is problematic.

A more recent proposal, Partial Information Decomposition (PID) \cite{WilliamsBeer2010}, explicitly aims to separate mutual information into unique, redundant, and synergistic information. However, because PID operates directly on observed probability distributions, its decomposition necessarily reflects both the functional relationships between inputs and output and the statistical structure of the input distribution. When inputs exhibit statistical dependencies—as commonly occurs in empirical data—these two sources of structure become inseparable, leading to fundamental ambiguity in attributing information to specific components. Multiple mutually incompatible axiomatic definitions have been proposed, each producing different decompositions from the same data~\cite{vanEnk2023, Kolchinsky2024, MedianoEtAl2022, MedianoEtAl2024, GomezFigueiredo2024}.

The implications of conflating functional and input distribution become clearer when we consider incomplete data. If observations cover only a subset of possible input combinations, the data provide an incomplete view of the system's input-output mapping. Different complete functions—each embodying a different information-processing architecture—may be consistent with the same partial observations. This underdetermination is not merely a practical limitation of sampling, but an inherent limitation: observations only determine the function where they sample. FID does not resolve the PID debate; it reframes the question by changing what is being decomposed.

To illustrate this underdetermination, consider two experimental devices, one implementing an OR function and the other an XOR function. An investigator collects data from each but, in both cases, observes only three of the four possible input patterns: \(f(0,0)=0\), \(f(0,1)=1\), and \(f(1,0)=1\), but never \(f(1,1)\). The two datasets are identical, yet the underlying systems are fundamentally different. From the data alone, what should the investigator infer about how the inputs contribute to the output?

This example illustrates that a partial dataset is like a shadow: it reveals some structure, but cannot uniquely identify the object that produced it. Here, the same data ``shadow'' is cast by two systems with distinct intrinsic functional architectures. The OR and XOR functions correspond to qualitatively different modes of information processing. In XOR, the output is determined only by the joint state of the inputs---pure synergy---whereas in OR, it is not. Partial observations cannot distinguish between these architectures.

Consequently, any dataset that does not exhaustively specify the system's input--output mapping cannot uniquely determine how information is distributed among the inputs. A rigorous framework must therefore define information decomposition with respect to the complete system and, when only a partial view is available, explicitly characterize the space of decompositions consistent with the observations.

Because FID analyzes functional structure directly, it requires specification of the output for every possible combination of input values— that is, the full Cartesian product of the input domains. Over this complete domain, inputs necessarily vary independently: each input exhibits all of its possible values in combination with all possible values of every other input. Under these conditions, inputs cannot share information with one another (i.e., have zero mutual information). As a result, the total mutual information thus partitions unambiguously into information attributable to individual inputs and synergistic information arising from their joint consideration(see section~\ref{FORMAL} for a full explanation).

When data do not support a complete functional specification, FID samples the space of complete functions consistent with the observations. Each sampled completion yields its own unique decomposition, and the resulting collection of decompositions reflects what the available data determine and what they leave uncertain. Thus, where observations fully specify the function, FID provides a single unambiguous decomposition; where they do not, FID characterizes not only the bounds on possible decompositions but also how different completions relate to one another. These relationships can be visualized to reveal structured patterns in the space of possible decompositions.

The reader may now be wondering, ``What of redundancy?'' Functional Information Decomposition operates in a regime where redundancy is not a feature of the decomposition by construction. Because FID is defined with respect to complete functional specification, informational contributions partition without remainder; no additional redundancy term is required or defined within the framework. In this context, redundancy does not arise, and decomposition concerns only independent and synergistic contributions. A full discussion of redundancy---including the distinction between operational (predictive) and informational (content-overlap) notions of redundancy---is beyond the scope of the present work; see~\cite{bohm2026what}.

To facilitate the practical use of FID, we provide a Python implementation as the open-source package \texttt{fid-tools}, available via:

PyPI (\texttt{pypi.org/project/fid-tools/}).

The package supports FID analysis for both complete and partially observed systems, enabling reproducible application and extension of the framework.

The remainder of this paper develops the formal apparatus underlying FID. We first establish the decomposition for complete functional specifications, and then extend it to incomplete cases via a principled sampling procedure that yields distributions over possible decompositions. Finally, we introduce an injectivity-based framing that characterizes coverage types according to how input mappings partition and overlap in the output space. This alternative perspective provides an independent structural interpretation of FID’s quantities, clarifying how different functional organizations give rise to independent versus synergistic information.

\section{Formal Overview}\label{FORMAL}

The mathematical basis of Functional Information Decomposition (FID) is defined with respect to a completely specified deterministic or probabilistic function \( f: X \to Y \).

\subsection{Complete Specification}
Let \( X_1, X_2, \ldots, X_n \) be discrete input variables with finite alphabets \(\mathcal{X}_i\), let \( X \) denote their joint, and let \( Y \) be a discrete output variable. 

\textit{Complete specification} holds when: 
\begin{enumerate}
    \item The mapping $P_{Y|X}$ is defined for every $\mathbf{x} \in \mathcal{X}_1 \times \cdots \times \mathcal{X}_n$, and
    \item The inputs are uniformly distributed over this product space, i.e., $P_X(\mathbf{x}) = 1/|\mathcal{X}|$.
\end{enumerate}

The uniformity in condition (2) is not an assumption about how the system is typically driven in practice. Rather, it defines a reference ensemble used to evaluate information intrinsic to the input–output mapping itself. This contrasts with input distributions arising under typical observational conditions, which are often strongly shaped by environmental, contextual, or experimental constraints and therefore represent the potentially arbitrary conditions under which the system was observed. Conditioning the decomposition on such distributions would entangle the system’s functional structure with contingent features of the world that supplies its inputs.

Intuitively, complete specification means that every possible input combination is represented, and the inputs vary independently across the full Cartesian domain.

\subsection{FID terms}

\textbf{Assuming complete specification, we define the following information quantities used by FID:}

\begin{itemize}
    \item \textbf{Total Information} (\(I_\text{tot}\))\textbf{.} The mutual information between the joined input set and the output:
\begin{equation}
    I_{\text{tot}} = I(Y; X)
\label{eq:total-info}
\end{equation}

\item \textbf{Synergistic Information} (\(I_\text{syn}(X \rightarrow Y\)))\textbf{.}
Synergy refers to the degree to which a whole is greater than the sum of its parts.
In an information-theoretic setting, this intuition can be formalized by comparing two
ways of describing the same set of inputs: as independent parts
\(X_1, X_2, \ldots, X_n\), or as a joint object \(X = (X_1,\ldots,X_n)\).

Under complete specification, where the inputs vary independently over their full
Cartesian product, these two perspectives are equivalent at the level of entropy:
\begin{equation}
H(X) = \sum_{i=1}^n H(X_i).
\label{eq:H_of_joint_vs_parts}
\end{equation}

In this respect, the whole and the collection of parts are equivalent: both carry the same uncertainty.

However, this equivalence does not generally persist under conditioning on an output \(Y\). Conditioning on the parts separately yields \(\sum_i H(X_i \mid Y)\), whereas conditioning the joint input yields \(H(X \mid Y)\). 
%Since these two quantities are equal before conditioning, any discrepancy that emerges must result from a decrease in uncertainty about \(Y\) when the inputs are considered jointly rather than separately (this can be equivalently viewed as the emergence of correlation among the inputs under conditioning on Y).
Since the input entropy itself is insensitive to joining, any discrepancy that appears after conditioning on \(Y\) reflects a difference in how much input entropy remains independent of \(Y\). Thus, if the conditional entropy is smaller when the inputs are considered jointly rather than one at a time, this must represent an increase in correlation with \(Y\). This phenomenon can be equivalently viewed—and is more typically described—as the emergence of correlations among the inputs under conditioning on \(Y\).

We therefore define the synergistic information as
\begin{equation}
I_{\text{syn}}(X \rightarrow Y)
=
\sum_{i=1}^n H(X_i \mid Y) - H(X \mid Y),
\label{eq:syn-info}
\end{equation}
which measures the degree to which the joint input is more informative about \(Y\) than the inputs treated independently. In this precise sense, synergistic information captures the intuition that the whole \(X\) is greater than the sum of its parts with respect to \(Y\).
Under this definition, \(I_{\text{syn}}(X \rightarrow Y) \ge 0\), with equality when the conditional entropy is insensitive to joining the inputs.

\item \textbf{Independent Information for the variable \( X_i \)} (\(I_\text{ind}(X_i)\))\textbf{.} The information about \( Y \) that is attributable uniquely to \( X_i \) considered in isolation.

Having defined synergy and substituting \(H(X_i|Y) = H(X_i) - I(X_i;Y)\) and \(H(X|Y) = H(X) - I(X;Y)\):

\begin{equation}
    I_\text{syn} = \sum_i H(X_i) - \sum_i I(X_i;Y) - H(X) + I(X;Y)
\label{eq:syn_ind_1}
\end{equation}

Under complete specification \(H(X) = \sum_i H(X_i)\), leaving:
\begin{equation}
    I(X;Y) = \sum_i I(X_i;Y) + I_\text{syn}
    \label{eq:syn_ind_2}
\end{equation}

Since each term is non-negative and their sum equals $I(X;Y)$ exactly, the information each input contributes independently is fully and exclusively captured by its individual relationship with $Y$. No portion of $I(X_i;Y)$ can be attributed to any other input without violating this equality. We therefore define:

\begin{equation}
    I_\text{ind}(X_i) = I(Y; X_i)
\label{eq:ind-info}
\end{equation}

With this definition for independent contribution we can alternativly define synergy as:
\begin{equation}
    I_\text{syn}(X \rightarrow Y) = I_\text{tot} - \sum_{i=1}^n I_\text{ind}(X_i)
\label{eq:syn_alt}
\end{equation}

\item \textbf{Solo-Synergy for variable \( X_i \)} (\(I_{\text{solo-syn}}(X_i)\))\textbf{.} The portion of total synergy that depends specifically on \( X_i \)’s participation. It is defined as the reduction in synergy when \( X_i \) is removed from the input set:
\begin{equation}
    I_{\text{solo-syn}}(X_i) = I_{\text{syn}}(X \rightarrow Y) - I_{\text{syn}}(X \setminus \{X_i\} \rightarrow Y)
\label{eq:solo-info}
\end{equation}
    or equivalently:
\begin{equation}
    I_{\text{solo-syn}}(X_i) = I(Y; X) - I(Y; X \setminus \{X_i\}) - I(Y; X_i)
\label{eq:solo-info-alt}
\end{equation}

\item \textbf{Bound on total solo-synergy.}
\begin{equation}
    2\, I_{\text{syn}}(X \rightarrow Y)
    \leq
    \sum_i I_{\text{solo-syn}}(X_i)
    \leq
    n\, I_{\text{syn}}(X \rightarrow Y)
\label{eq:solo-bounds}
\end{equation}
Here, $n$ is the number of input variables. This bound reflects the joint nature of synergy: at least two inputs must contribute for information to be synergistic (hence the lower bound of \(2 \times I_{\text{syn}})\). Technically, the lower bound emerges because synergistic information must be jointly supported: removing any single participating input eliminates at least part of the synergy, and each unit of synergy must be ‘counted’ in the solo-synergy of at least two inputs. At the other extreme, every input may participate fully in the synergy (yielding the upper bound of \(n \times I_{\text{syn}}\)). Since synergy is a joint property, excluding a synergistic input from the input set necessarily diminishes the synergistic potential of the remaining inputs. Moreover, any synergistic information loss from removing one input will be reflected in the solo-synergy of the inputs that participated in that synergy.

\item \textbf{Information Loss for variable \( X_i \)} (\(I_{\text{loss}}(X_i)\))\textbf{.} The total reduction in mutual information between \(X \) and \( Y \) when \( X_i \) is excluded from the input set:
\begin{equation}
    I_{\text{loss}}(X_i) = I_{\text{ind}}(X_i) + I_{\text{solo-syn}}(X_i)
\label{eq:loss}
\end{equation}
\end{itemize}

\subsection{Interpretation and Consequences}
In summary, the total information between the inputs \( X \) and the output \( Y \) decomposes as:
\begin{equation}
I_{\text{tot}} = \sum_{i=1}^n I_\text{ind}(X_i) + I_{\text{syn}}(X \rightarrow Y)
\label{eq:decomp}
\end{equation}

Although algebraically similar expressions appear in the literature, prior iterations of this algebraic form lack the complete specification requirement that guarantees non-negativity and uniqueness.

\vspace{1em}
Finally, because the inputs are statistically independent in FID, any subset of variables can be grouped into a single composite input without altering the independent or solo-synergistic contributions of the remaining variables.

For example, if \( X = \{X_1, X_2, X_3\} \), we may define a transformed input set \( X' = \{X_1, q\} \) where \( q = (X_2, X_3) \). The values of \( I_{\text{ind}}(X_1) \) and \( I_{\text{solo-syn}}(X_1) \) are invariant under this transformation.  
This property enables focused analysis of individual inputs while treating the remainder of the system as a unified block, without altering the decomposition results for the inputs of interest.

\vspace{0.5em}

These definitions provide an exact and unambiguous decomposition for any completely specified function. The following section demonstrates this fact with canonical examples. However, in practical applications, one rarely has access to the complete function. Section 4, therefore, addresses this by generalizing FID: we sample the space of complete functions consistent with partial observations. The FID terms calculated from these samples provide robust bounds, and their visualization reveals the detailed structure of the solution space.

\section{FID examples using complete functions}\label{full_spec_examples}

The following examples start with simple deterministic functions and progress to higher-dimensional and finally probabilistic functions. Each example is chosen to illustrate the range of behaviors captured by FID -- particularly how synergistic, independent, and solo-synergistic contributions manifest across different functional architectures.

\subsection{Example: AND}

\noindent
\textbf{Truth table:}
\[
\begin{array}{ccc}
X_1 & X_2 & Y=f(X_1,X_2) \\
\hline
0 & 0 & 0 \\
0 & 1 & 0 \\
1 & 0 & 0 \\
1 & 1 & 1
\end{array}
\]

\noindent
\textbf{FID results:}
\[
\begin{array}{lcc}
\text{Total Information:} & 0.811~\text{bits} & \\[4pt]
\text{Synergy:} & 0.189~\text{bits} & \\[2pt]
\hline
\text{Input} & \text{Independent} & \text{Solo-Synergy} \\
\hline
X_1 & 0.311~\text{bits} & 0.189~\text{bits} \\
X_2 & 0.311~\text{bits} & 0.189~\text{bits} \\
%\hline
\end{array}
\]

Because there are only two inputs, and synergy, by definition, must be located in more than one, the synergy depends on both.

\subsection{Example: XOR}

\noindent
\textbf{Truth table:}
\[
\begin{array}{ccc}
X_1 & X_2 & Y=f(X_1,X_2) \\
\hline
0 & 0 & 0 \\
0 & 1 & 1 \\
1 & 0 & 1 \\
1 & 1 & 0
\end{array}
\]

\noindent
\textbf{FID results:}
\[
\begin{array}{lcc}
\text{Total Information:} & 1.000~\text{bits} & \\[2pt]
\text{Synergy:} & 1.000~\text{bits} & \\[4pt]
\hline
\text{Input} & \text{Independent} & \text{Solo-Synergy} \\
\hline
X_1 & 0.000~\text{bits} & 1.000~\text{bits} \\
X_2 & 0.000~\text{bits} & 1.000~\text{bits} \\
%\hline
\end{array}
\]

Because neither input provides any information about $Y$ individually, all mutual information is synergistic. The XOR function represents \textit{pure synergy}.

\subsection{Example: LED Square} \label{LED_square}

\includegraphics[width=6cm]{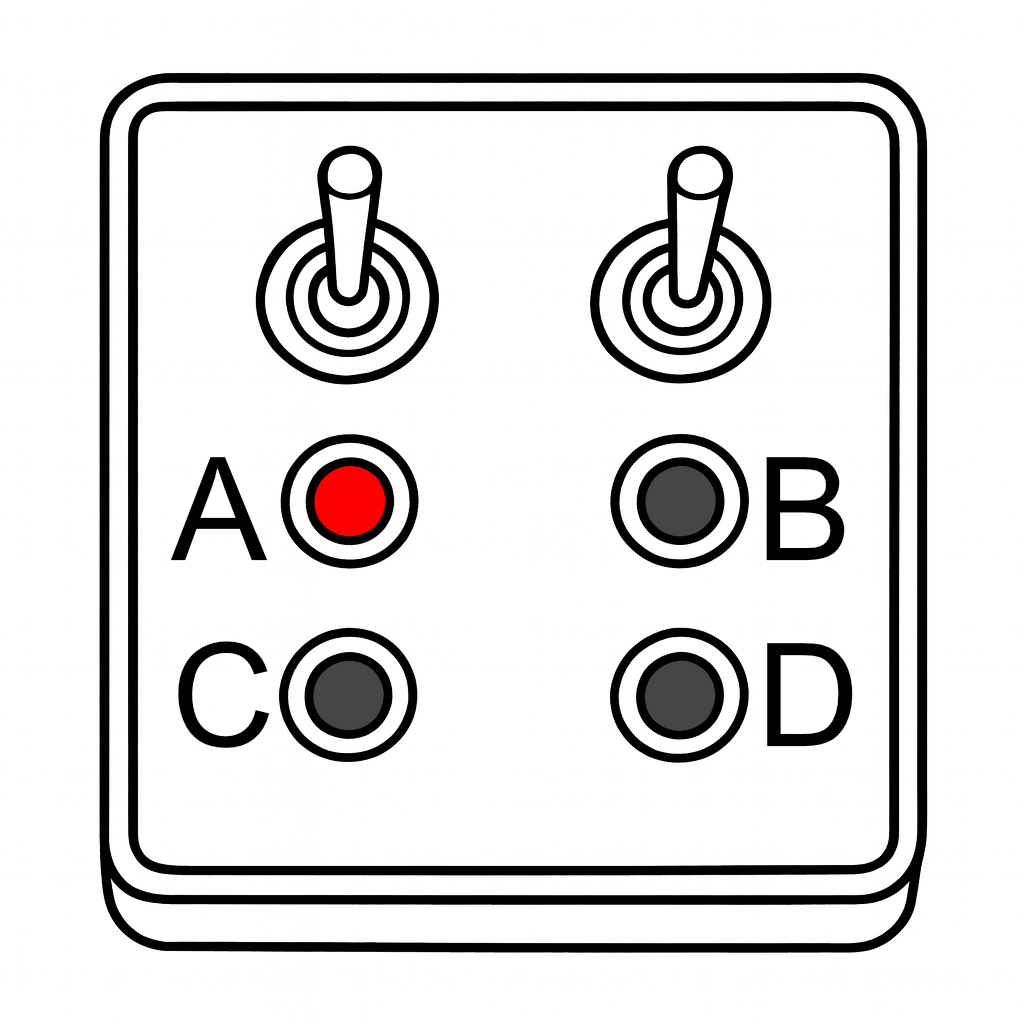}

Consider a simple device with two toggle switches and four LEDs arranged in a square. 
Each switch ($s_1$ and $s_2$) can be either up ($U$) or down ($D$), and together they determine which of four LEDs ($A$, $B$, $C$, $D$) is lit. 
This mapping can be expressed as the function $f(s_1, s_2) \to Y$ shown below.

\[
\begin{array}{ccc}
s_1 & s_2 & Y=f(s_1,s_2) \\
\hline
U & U & A \\
U & D & B \\
D & U & C \\
D & D & D
\end{array}
\]

\noindent
\textbf{FID results:}
\[
\begin{array}{lcc}
\text{Total Information:} & 2.000~\text{bits} & \\[2pt]
\text{Synergy:} & 0.000~\text{bits} & \\[4pt]
\hline
\text{Input} & \text{Independent} & \text{Solo-Synergy} \\
\hline
s_1 & 1.000~\text{bits} & 0.000~\text{bits} \\
s_2 & 1.000~\text{bits} & 0.000~\text{bits} \\
%\hline
\end{array}
\]

This example illustrates that a system can depend on multiple inputs without exhibiting synergy. 
Here, $s_1$ independently determines whether the lit LED is in the top or bottom row ($A,B$ vs.\ $C,D$), while $s_2$ independently determines whether it is on the left or right column ($A,C$ vs.\ $B,D$). 
Together, they specify the output uniquely, but without shared or synergistic information.

\subsection{Example: Game of Life}

Conway’s Game of Life (GoL) is a zero-player cellular automaton introduced by John Conway in 1970. It is played on an infinite grid of binary-valued cells (alive or dead), evolving in discrete time steps according to three simple rules:
\begin{enumerate}
    \item Any dead cell with exactly three live neighbors becomes alive.
    \item A live cell with two or three live neighbors survives.
    \item All other cells become dead.
\end{enumerate}

GoL is a canonical example of how simple local rules can generate complex, emergent behavior.

\includegraphics[width=8cm]{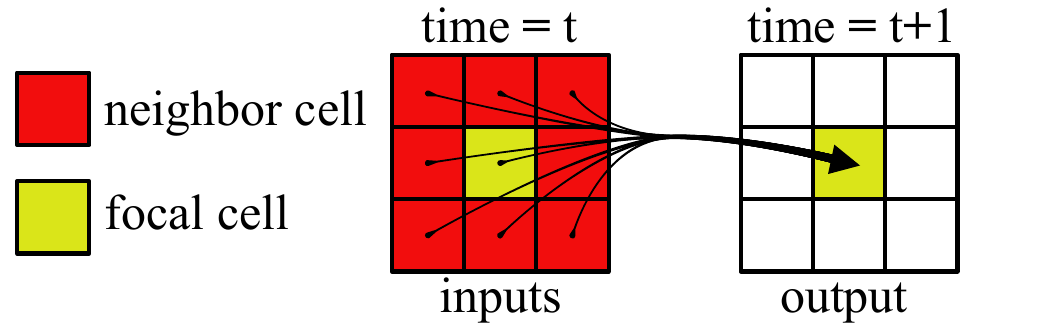}

Here, FID is applied to assess how the next state of a focal cell depends on the prior states of its $3\times3$ neighborhood. Each location in the $3\times3$ grid represents an input at time $t$, and the value of the center cell at time $t{+}1$ is the output. There are nine binary inputs, producing $2^9 = 512$ possible input combinations, each consisting of a 9-bit vector.
\vspace{1em}

\noindent
\textbf{FID results:}
\[
\begin{array}{lcc}
\text{Total Information:} & 0.84635~\text{bits} & \\[2pt]
\text{Synergy:} & 0.63765~\text{bits} & \\[4pt]
\hline
\text{Input} & \text{Independent} & \text{Solo-Synergy} \\
\hline
\text{Neighbor cells (×8)} & 0.02472~\text{bits each} & 0.358~\text{bits each} \\
\text{Center cell} & 0.01092~\text{bits} & 0.099~\text{bits} \\
%\hline
\end{array}
\]

The total synergy of $0.63765$ bits indicates that the future state of a cell depends predominantly on joint configurations of multiple neighbors rather than on any single cell alone. The solo-synergy values show that neighboring cells are approximately $3.5\times$ more predictive of the next state than the focal cell itself.

\subsection{Example: Probabilistic Function}
The previous examples demonstrated the decomposition for simple, well-known logical operations. We now examine a more complex probabilistic function, where the provided inputs do not fully account for the entropy of the output. This system consists of three binary inputs $X_1, X_2, X_3$ and a discrete output $Y \in \{0,1,2,3\}$, with the mapping:

\vspace{0.5em}

\renewcommand{\arraystretch}{1.25}
\[
\begin{array}{ccc|cccc}
X_1 & X_2 & X_3 & P(Y{=}0) & P(Y{=}1) & P(Y{=}2) & P(Y{=}3) \\[0pt]
\hline
%  &   &   & \\[0pt]
0 & 0 & 0 & \frac{1}{4} & \frac{1}{4} & \frac{1}{4} & \frac{1}{4} \\[0pt]
0 & 0 & 1 & \frac{1}{20} & \frac{9}{20} & 0 & \frac{1}{2} \\[0pt]
0 & 1 & 0 & \frac{3}{10} & \frac{1}{5} & 0 & \frac{1}{2} \\[0pt]
0 & 1 & 1 & 0 & 0 & 1 & 0 \\[0pt]
1 & 0 & 0 & 1 & 0 & 0 & 0 \\[0pt]
1 & 0 & 1 & 0 & 1 & 0 & 0 \\[0pt]
1 & 1 & 0 & 1 & 0 & 0 & 0 \\[0pt]
1 & 1 & 1 & 0 & 0 & 0 & 1 \\[0pt]
\end{array}
\]

\noindent
\textbf{FID results:}
\[
\begin{array}{lcc}
\text{Total Information:} & 1.3627~\text{bits} & \\[2pt]
\text{Synergy:} & 0.5463~\text{bits} & \\[4pt]
\hline
\text{Input} & \text{Independent} & \text{Solo-Synergy} \\
\hline
X_1 & 0.2309~\text{bits} & 0.3676~\text{bits} \\
X_2 & 0.1886~\text{bits} & 0.3512~\text{bits} \\
X_3 & 0.3968~\text{bits} & 0.4105~\text{bits} \\
\hline
\end{array}
\]

Because the function is probabilistic, the inputs do not fully determine the output ($H(Y|X) = H(Y) - I(X;Y)$ defines the residual entropy in the output that is not dependent on the inputs). Here, the output has $H(Y)=1.9527$ bits of entropy, while the inputs account for only $I(X;Y)=1.3627$ bits. The remaining $0.5900$ bits reflect variability in $Y$ that the inputs cannot explain. This residual uncertainty may arise from an unobserved input or from intrinsic randomness, but in either case it is not part of $I(X;Y)$ and therefore does not appear in the decomposition.

\section{Completing Partial Functions}

The decomposition presented in Section 2 is defined for complete functions—mappings where every possible input combination has a specified output and all inputs are independent. When data leave input patterns unobserved, the function is underdetermined, and the formal apparatus cannot be directly applied. Incompleteness arises from two distinct sources, each requiring a different procedure to restore the conditions under which FID operates:

\subsection{Handling Dependent Inputs}\label{section:handling-dependent-inputs}
One source of incompleteness arises from input dependencies that are not due to incomplete observations, but to model misspecification. For predictive modeling, a misspecified input space may be harmless. If unreachable states are never encountered, predictions remain accurate, and the model works. A model whose only goal is accurate prediction can carry extra degrees of freedom without consequence. However, when a model includes impossible inputs, it cannot be represented as a completely specified function. The inputs are not simply ``missing'' but undefined.

Consider an animal shelter tracking system with inputs: Dog (Y/N), Cat (Y/N), Other (Y/N), and Sex (M/F). While this will capture all relevant information, most input combinations are impossible. For example, an animal cannot simultaneously be classified as a dog and a cat. Treating Dog, Cat, and Other as independent inputs misrepresents the system's capacity: it suggests a function with 16 possible input states across four independent binary variables when only 6 of these states are reachable. When FID is applied, the misspecified input space leads to a decomposition based on that inflated space, thereby characterizing a system that does not exist. In this case, the inputs should be recast as animal and sex. This will limit the input combinations to the six possible pairs: dog+male, dog+female, cat+male, cat+female, other+male, and other+female. 

On the other hand, missing input patterns in observed data do not always indicate structural impossibility. Absence from the dataset may, in fact, reflect incomplete sampling rather than an inherent constraint. Unfortunately, determining whether unobserved patterns are truly unreachable, and thus require recasting the input space, is not possible from observations alone. The decision to merge inputs is a modeling choice that requires domain knowledge or other priors about the system's true structure; choices that must be resolved before FID can be applied.

\subsection{Handling Missing Observations}\label{section:handling-missing-observations}

Once the input space correctly reflects the system's degrees of freedom, the second source of incompleteness can be addressed: unobserved input patterns that leave the function underspecified. FID addresses this by systematically sampling the completion space. We construct a set of sample functions by assigning output values to every unobserved input pattern, producing complete functions compatible with observations. The values used in these constructions may represent an exhaustive enumeration, or,  in cases where exhaustive enumeration is infeasible --- either because the number of possible completions is too large or because the output space is probabilistic --- the points may represent a partial enumeration obtained via random sampling. \footnote{When random sampling is required, FID does not define the sampling process. We have had success with Dirichlet sampling, which provides good coverage for a given number of samples. This type of sampling includes a concentration parameter that controls the degree of probabilism in the sampled completions, ranging from near-deterministic to maximally probabilistic. This tuning ensures that we cover the range of possible values while also providing useful structural information.}

The resulting set of functions represents the space of functions that the data permit.
FID does not approximate a single ``true'' function—it characterizes the landscape of functions consistent with observations. Referring back to the metaphor that partial data is like a shadow, the set of completion functions represents all possible objects that could generate a given shadow. Each sample is analyzed using the definitions in section~\ref{FORMAL}, yielding an exact decomposition. The ensemble reveals both the bounds allowed by the data and the distribution of decomposed information, making uncertainty in the decomposition explicit and quantifiable.

\section{FID examples using incomplete functions}
In this section, we provide examples of FID results from incomplete data. We start with a formal presentation of the OR versus XOR thought exercise presented in the introduction.

\subsection{Incomplete Function: OR--XOR Continuum}\label{EXAMPLE_OR_XOR}

Consider \(f(X_1,X_2)\to Y\) with one unobserved input–output pair:
\[
\begin{array}{ccc}
X_1 & X_2 & Y=f(X_1,X_2) \\
\hline
0 & 0 & 0 \\
0 & 1 & 1 \\
1 & 0 & 1 \\
1 & 1 & \text{unknown}
\end{array}
\]
If \(f(1,1)=1\), the function is OR; if \(f(1,1)=0\), it is XOR.  
With incomplete specification, the value of \(Y\) for \(f(1,1)\) is unconstrained: it could be \(0\), \(1\), or any probabilistic mixture of the two—for example, uniform (\(50\%\) 0, \(50\%\) 1) or skewed (\(10\%\) 0, \(90\%\) 1).  
The missing entry, therefore, allows us to define a \emph{space} of possible completions, ranging continuously from OR to XOR and including all probabilistic variants between them.

\vspace{1em}

\noindent\textbf{FID results from sampled completions} (all values in bits):
\[
\begin{array}{lcc}
%\textbf{Measure} & \textbf{Minimum (bits)} & \textbf{Maximum (bits)} \\
%\hline
\text{Total Information} & 0.6887 - 1.0000 \\
\text{Synergy} & 0.1887 - 1.0000 \\
\hline
\text{Input} & \text{Independent} & \text{Solo-Synergy} \\
\hline
X_1 & 0.0000 - 0.3113 & 0.1887 - 1.0000 \\
X_2 & 0.0000 - 0.3113 & 0.1887 - 1.0000 \\
%\hline
\end{array}
\]

\begin{figure}[h]
    \centering
    \includegraphics[width=.5\textwidth]{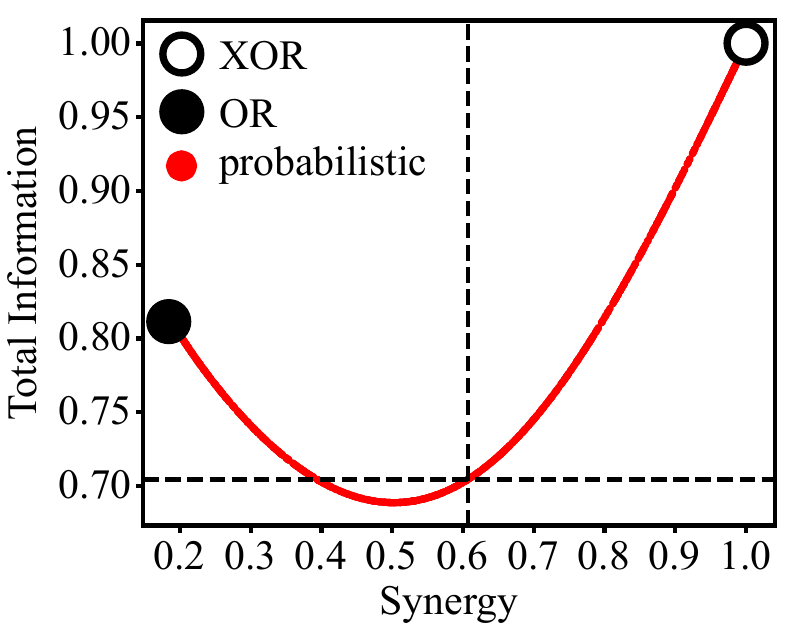}
    \caption{\textbf{OR--XOR completion space.}
    The total information and synergy for completions for a partially observed function compatible with OR or XOR. Black markers indicate deterministic completions (solid dot: OR; open circle: XOR). Red indicates probabilistic completions.
    The dashed lines intersect at the maximum-entropy (“neutral”) completion, where the missing input condition $X_1 = 1, X_2 = 1$, is equally likely to yield $Y = 0$ or $Y = 1$.
    }
    \label{fig:OR_XOR}
\end{figure}

While the bounds provide helpful information, visualizations reveal far richer insights. 
Figure~\ref{fig:OR_XOR} illustrates the solution space. 
The black dot and black circle mark the deterministic completions—OR and XOR, respectively. 
The red points show FID values associated with probabilistic completions. 
The dashed lines intersect at the maximum-entropy case, corresponding to the solution that introduces the fewest assumptions. 

Figure~\ref{fig:OR_XOR} illustrates how FID is not tied to a single outcome but characterizes the entire landscape of informational decompositions permitted by the incomplete specification.
This range of values is not uncertainty about a single true value, but rather a direct reflection of the fact that the underlying function is underdetermined by the collected data. Each possible completion of the missing input-output pair defines a different, equally valid function, each with its own unique informational structure.

\subsection{Three input function}\label{Three_input_function}

In this example, we consider the function $Y =$ IF $X_1$ THEN $X_2$ ELSE $X_3-X_2$ first as completely specified and then when some of the function is hidden. The function can alternatively be expressed as,
\begin{equation}
Y = \begin{cases} X_2 & \text{if } X_1 \\      X_3 - X_2 & \text{otherwise}, \end{cases}
\label{eq:ThreeInputExample}
\end{equation}
or in the form of a lookup table:

\[
\begin{array}{cccc}
X_1 & X_2 & X_3 & Y=f(X_1,X_2,X_3) \\
\hline
0 & 0 & 0 & 0\\
0 & 0 & 1 & 1\\
0 & 1 & 0 & -1\\
0 & 1 & 1 & 0\\
1 & 0 & 0 & 0\\
1 & 0 & 1 & 0\\
1 & 1 & 0 & 1\\
1 & 1 & 1 & 1\\
\end{array}
\]

The completely specified function yields the unique decoposition:
\[
\begin{array}{lcc}
\text{Total Information:} & 1.4056~\text{bits} & \\[2pt]
\text{Synergy:} & 0.8444~\text{bits} & \\[4pt]
\hline
\text{Input} & \text{Independent} & \text{Solo-Synergy} \\
\hline
X_1 & 0.1556~\text{bits} & 0.5944~\text{bits} \\
X_2 & 0.2500~\text{bits} & 0.7500~\text{bits} \\
X_3 & 0.1556~\text{bits} & 0.3444~\text{bits} \\
\hline
\end{array}
\]

This indicates that $X_2$ carries the most information, both independently and synergistically, of the three inputs, which is expected since $X_1$ only acts as a switch and $X_3$ only participates in one of the conditions determined by $X_1$.

\begin{figure}[H]
    \centering
    \includegraphics[width=1\textwidth]{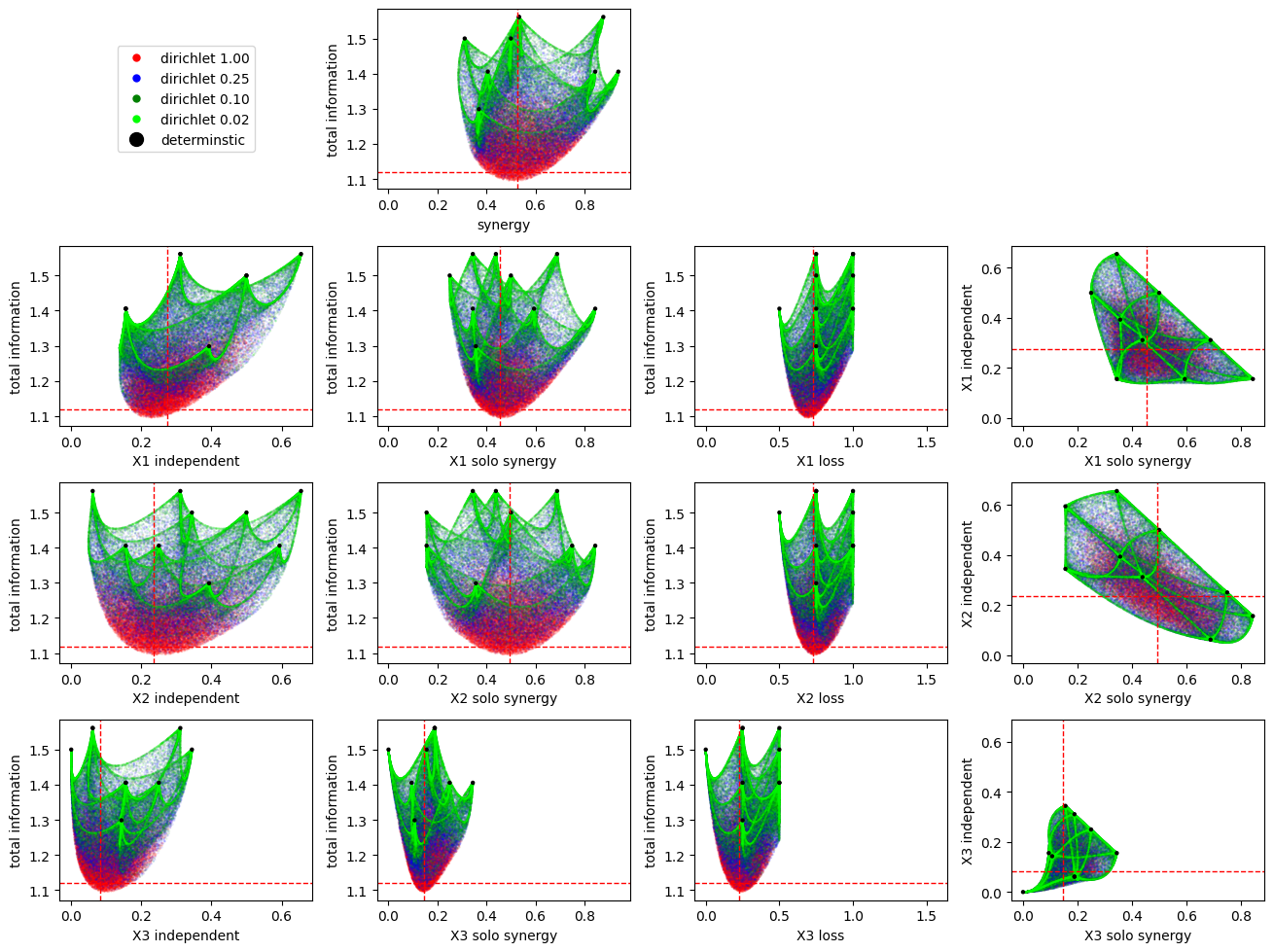}
    \caption{FID decomposition for the function defined in equation~\ref{eq:ThreeInputExample} if when two inputs $(0,0,0)$ and $(0,1,1)$ are unobserved. Each panel shows decompositions for functions consistent with the observed data, obtained by sampling. Black points indicate decompositions resulting from fully deterministic completions. Colored points correspond to decompositions of probabilistic completions, with the color indicating the Dirichlet concentration parameter (green: more deterministic; red: more probabilistic). Rows correspond to individual inputs, and columns show different projections for each input, as labeled. Dashed lines indicate values obtained from the maximum entropy assumption.}
    \label{fig:ThreeInputExample}
\end{figure}

Now consider that we have collected data consistent with the function described in equation~\ref{eq:ThreeInputExample}, except that we have not observed the input conditions $(X_1,X_2,X_3)$ = $(0,0,0)$ or $(0,1,1)$. In figure~\ref{fig:ThreeInputExample}, we consider the space of decompositions compatible with the partial observation. Because there are two unspecified outputs, each of which can take three values $\{-1,0,1\}$, there are nine deterministic completions consistent with the data. These deterministic completions are indicated with black points. The probabilistic completions are indicated using several colors, each corresponding to a different Dirichlet concentration. The missing data result in a partially defined function, in which the relative roles of X1 and X2 are less clear than they appear under complete specification. However, the missing data do not obscure the fact that $X_3$ clearly still plays a smaller role than either of the other two inputs. 

\subsection{Prediction from Genetic Data}\label{GenExample}

\begin{figure}[!h]
    \centering
    \includegraphics[width=1\textwidth]{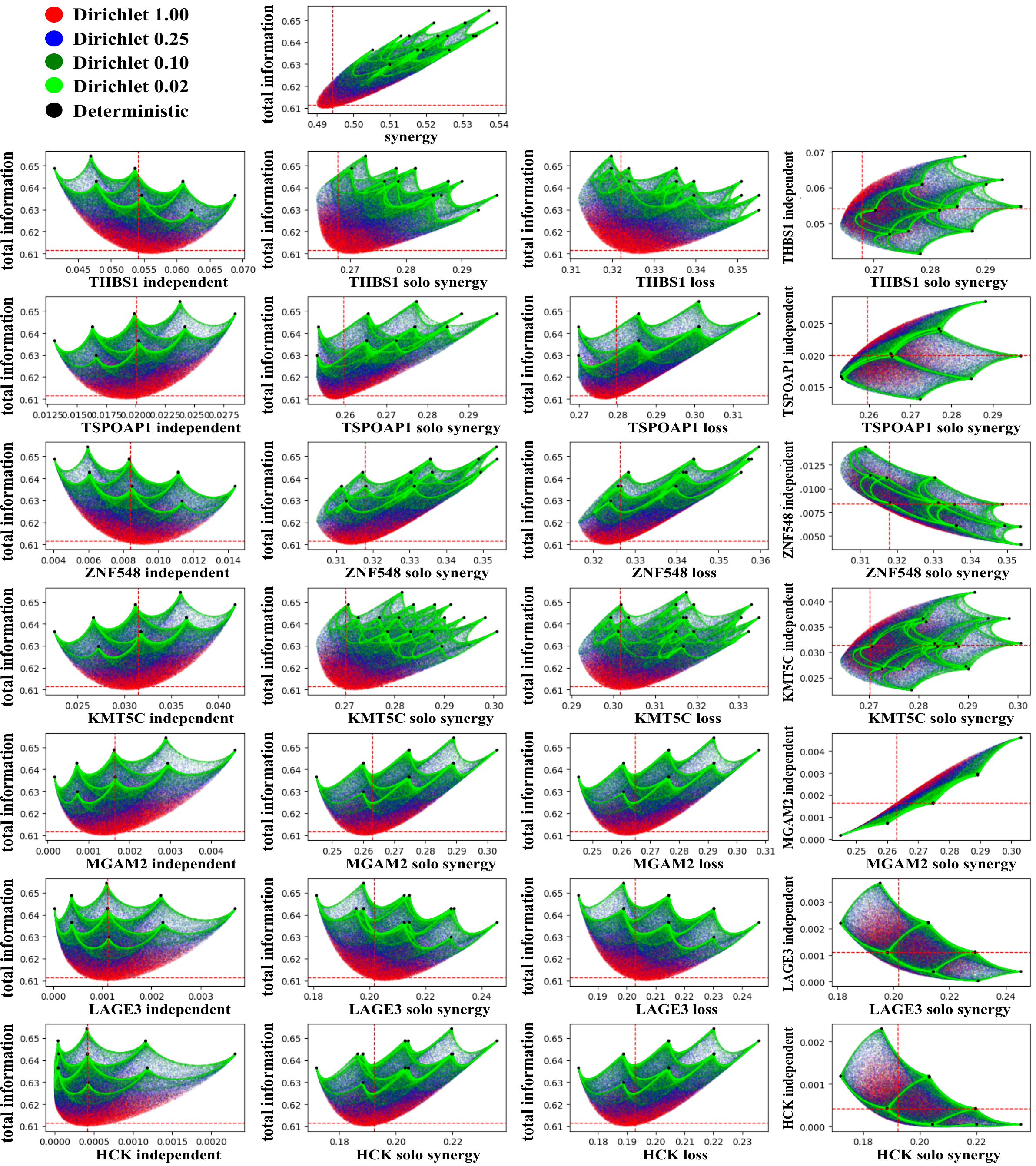}
    \caption{FID decomposition for a seven-gene predictor of resistance to the apoptosis inhibitor ABT737. Each panel shows decompositions for functions consistent with the observed genetic data, obtained by sampling. Black points indicate decompositions resulting from fully deterministic completions. Colored points correspond to decompositions of probabilistic completions, with the color indicating the Dirichlet concentration parameter (green: more deterministic; red: more probabilistic). Rows correspond to individual genes, and columns show different projections for each gene, as labeled. Dashed lines indicate values obtained from the maximum entropy assumption.}
    \label{fig:gene_data}
\end{figure}

In figure~\ref{fig:gene_data}, we decompose a function that indicates whether patients with a particular gene expression profile are susceptible or resistant to a particular anti-cancer drug. The list of genes/transcripts that are informative about resistance to the drug ABT737 (an apoptosis inhibitor) was obtained by maximizing the amount of information that sets of genes have about the binary variable resistant/susceptible, using susceptibility data collected from the ``Genomics of Drug Sensitivity in Cancer" project~\cite{Yangetal2013} in 536 cell lines, and transcriptomics for those same cell lines obtained from the DepMap portal~\cite{Arafehetal2025}. Expression levels of transcripts\footnote{The DepMap portal lists expression data in terms of TPM (transcripts per million) for 57,975 RNA sequences obtained in RNASeq experiments. These mRNA sequences correspond to genes in open reading frames, alternative transcripts, pseudogenes, as well as non-coding RNAs.} are binarized using a maximum entropy procedure creating the expression profile~\cite{Kubaletal2026}. The information-theoretic search identified the seven transcripts (``genes" from here-on out) that have (as a group) the most information about the resistance variable. This data set is interesting from the point of view of information decomposition because the set is highly synergistic(see~\cite{Kubaletal2026} for more details).

We evaluated the seven genes with the highest mutual information with the class variable. Even after preprocessing, there are $2^7$ possible deterministic input patterns. The observed data (536 records of the state of 7 genes as on/off, and the IC50 of drug resistance converted to a binary variable via $Z$ transformation) represent deterministic input–output pairs. Within the data, all but four input patterns were present. To generate the point clouds, 400,016 samples were assessed. Sixteen samples were the complete set of deterministic solutions. The remaining samples were probabilistic solutions generated using Dirichlet sampling of the 16-dimensional space, with 100,000 samples at each of four different Dirichlet alpha values. Green points correspond to more deterministic completions generated with low Dirichlet alpha, while red points correspond to more probabilistic completions generated with high alpha. Each plot displays a distinct projection of a high-dimensional manifold, offering insight not only into the bounds on the space of solutions but also into its topology. The third column shows loss, indicating how much total predictive power is dependent on each gene, while the first two columns decompose this loss into independent and solo-synergy terms.
While numerical ranges could be tabulated, the visualization conveys the structure of the data far more effectively, capturing relationships that scalar summaries cannot.
These results delineate the range of information architectures consistent with the observed data, enabling principled inferences about genetic contribution and dependence while maintaining explicit awareness of uncertainty.

\section{Multi-Output Systems and Output Scope}\label{sec:multi-output}

Many real systems produce multiple outputs. FID accommodates this naturally, without modification to its core formalism.

A system with multiple outputs may be represented as a function
\begin{equation}
f : X \rightarrow (Y_1, Y_2, \dots, Y_m),
\label{eq:multi-out}
\end{equation}
where $X = \{X_1, X_2, \dots\}$ denotes the set of inputs. Since each output defines a distinct function of the same inputs, the formalism developed in the previous sections can be applied to each of these output-specific functions. 

In addition, any subset of outputs may be combined into a single composite variable with a larger alphabet, and FID can be applied to the resulting mappings. Each such analysis is well-defined and answers a distinct question about how the inputs determine a particular aspect of the system's output. When outputs are considered individually, each decomposition captures how inputs determine that particular output component. When outputs are aggregated, new structure may appear: dependencies between output components can introduce synergistic or independent contributions that are absent from any single-output decomposition.

A systematic theory of output-scope structure---formalizing how decompositions relate across output groupings---is deferred to future work, but the basic intuition is clear: by examining how FID's quantities change across output groupings, one can identify which input-output relationships are visible only at particular output resolutions and whether those relationships involve individual inputs or joint input configurations.

\section{Coverage Type: A Geometric Perspective}

Functional Information Decomposition (FID) quantifies how information is distributed among inputs: how much is attributable to individual inputs, and how much emerges only from their joint consideration. But what structural property of a function determines this distribution?

The answer is geometric. Each value within the domain of each input maps to some region of the output space---a single point in the deterministic case or a set of possible outputs in the stochastic case. When these regions are disjoint, observing the output reveals which input value (of that variable) produced it. In these cases, the input contributes information independently. When regions overlap, the same output can be produced by different input values. This results in structural ambiguity. When this ambiguity can only be resolved by considering other similarly ambiguous inputs, synergy emerges.

This picture is immediately visible in the examples from Section~\ref{full_spec_examples}. In the LED square, each switch state maps to a distinct pair of outputs; the regions are disjoint, and FID finds no synergy. In XOR, both input values map to the same output set $\{0,1\}$; the regions overlap completely, and FID finds pure synergy. OR is intermediate; some input states map uniquely, while others overlap, producing a mixture of independent and synergistic information.

To formalize this intuition, we introduce a classification of functions by their \textbf{coverage type}---the pattern by which input supports partition or overlap in the output space. This framework extends the classical mathematical concept of injectivity to stochastic functions, describing the same phenomena that FID quantifies but in structural, set-theoretic terms\footnote{If the two frameworks appear to disagree, the information-theoretic decomposition provides the definitive quantitative assessment.}. Coverage type provides geometric intuition for FID's quantities, suggests alternative proof strategies, and builds on different mathematical machinery that may enable extensions inaccessible from the information-theoretic viewpoint alone.

We begin by extending the standard mathematical definition of injectivity to characterize the spectrum of functional relationships.

\subsection{Injective, Pseudo-Injective, and Overjective}
\begin{figure}[h]
\centering
\includegraphics[width=0.9\textwidth]{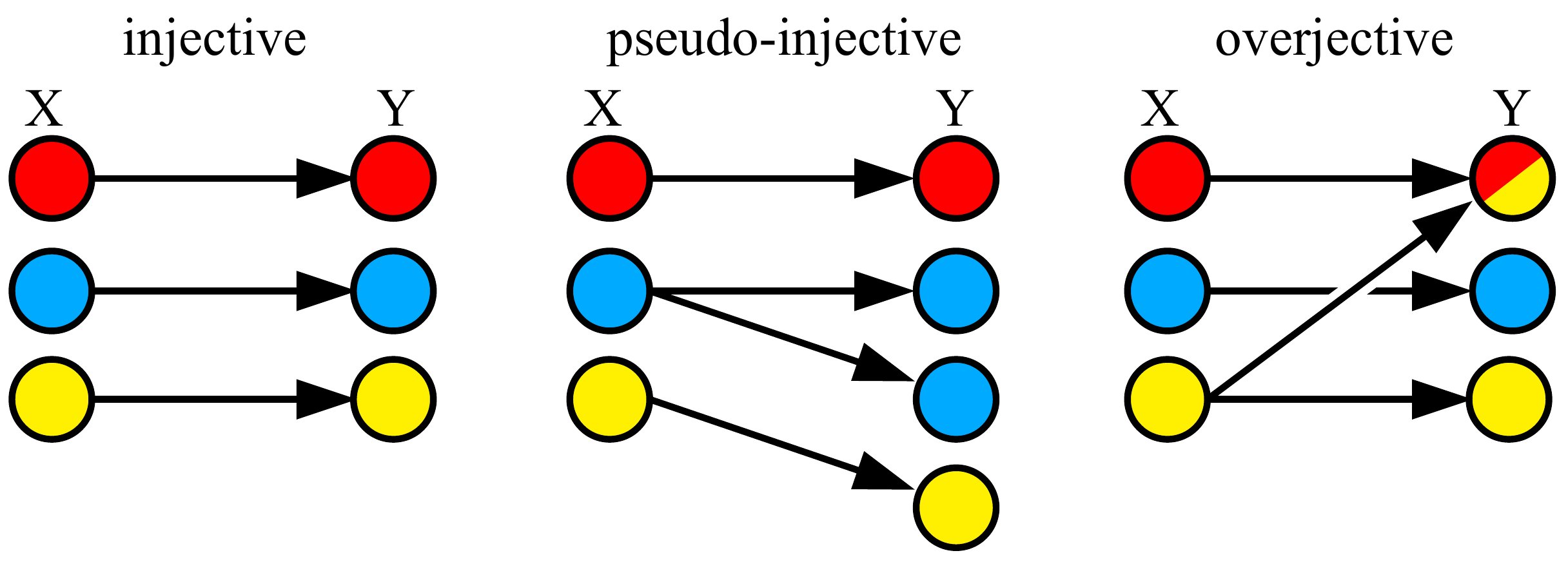}
\caption{Coverage types illustrated. \textbf{Left (Injective):} Each input value (X) maps to a unique output value (Y). \textbf{Center (Pseudo-injective):} Each input value maps to a distinct, non-overlapping subset of outputs—supports remain disjoint. \textbf{Right (Overjective):} Input supports overlap; the split-colored (red/yellow) output indicates that multiple input values can produce the same output, creating fundamental ambiguity.}
\label{fig:coverage_types}
\end{figure}
The classical notion of injectivity stems from the Latin \textit{jacere}---``to throw.'' An injective function maps each input to a unique output. We extend this to stochastic functions based on the support of an input value: the set of outputs it can produce. When supports are disjoint, each input is identifiable from the output regardless of stochasticity. When supports overlap, structural ambiguity arises because the output no longer identifies the input. The relationship among these supports is what we call the coverage type and is illustrated in figure~\ref{fig:coverage_types}.

\begin{itemize}
    \item \textbf{Injective coverage}: Each input maps to a unique, single output---the classical case.
    
    \item \textbf{Pseudo-injective coverage}: Each input scatters across a region, but these regions remain distinct and non-overlapping. Different inputs cover different territories.
    
    \item \textbf{Overjective coverage}: Input supports overlap. Multiple inputs can produce the same outputs.
\end{itemize}

We now formalize these definitions. In standard set theory, a function $f: X \to Y$ is \textbf{injective} (one-to-one) if distinct input values map to distinct output values: $f(x_1) = f(x_2)$ implies $x_1 = x_2$. This ensures that each input value specifies a unique output value without ambiguity.

For the general case, each input $x \in X$ corresponds to its \emph{output support}
\begin{equation}
S_x = \{ y \in Y \mid P(Y=y \mid X=x) > 0 \}.
\label{eq:support}
\end{equation}
The coverage type is then determined by the relationship among these supports:

\begin{itemize}
    \item \textbf{Injective:}
\begin{equation}
    \forall x_1 \neq x_2, \; S_{x_1} \cap S_{x_2} = \varnothing
    \quad \text{and} \quad |S_x| = 1.
\label{eq:injective}
\end{equation}
    A function is \textbf{injective} if each input value maps to a unique output value. This represents a state of perfect, unambiguous determination. This is the accepted definition in mathematics. \emph{Note: In the context of FID, where $Y$ is defined as the set of observed outputs, injective functions are necessarily bijective.}
    
    \item \textbf{Pseudo-Injective:}
\begin{equation}
    \forall x_1 \neq x_2, \; S_{x_1} \cap S_{x_2} = \varnothing,
    \quad \text{and} \quad \exists x \text{ such that } |S_x| > 1.
\label{eq:pseudo-injective}
\end{equation}
    A function is \textbf{pseudo-injective} if distinct input values map to \emph{disjoint, non-overlapping subsets} of output values. While a single input value may not uniquely determine a single output value, it restricts the output to a specific, well-defined subset. The ``meaning'' of an input value, defined by the subset of output values it permits, is fixed and independent.
    
    \item \textbf{Overjective:}
\begin{equation}
    \exists x_1 \neq x_2 \text{ such that } S_{x_1} \cap S_{x_2} \neq \varnothing.
\label{eq:overjective}
\end{equation}
    A function is \textbf{overjective} if distinct input values map to \emph{overlapping} subsets of output values. This overlap introduces fundamental ambiguity: knowledge of the output value alone is insufficient to determine the specific input value that produced it, as multiple input values could be responsible.
    In the deterministic limit, this reduces to the many-to-one case where
    $f(x_1)=f(x_2)$ for some $x_1 \neq x_2$.
\end{itemize}

\subsubsection{Coverage Type and Information Decomposition}

Each coverage pattern relates directly to FID's decomposition of information.

\textbf{Injective coverage} corresponds directly and trivially to independent information. Each input value uniquely determines the output—no ambiguity exists.

\textbf{Pseudo-injective coverage} also yields independent information. Each input constrains the output to a disjoint subset. Even when multiple inputs are needed to fully determine the output, their contributions remain independent because their constraint sets don't overlap. The LED square example demonstrates this: each switch eliminates half the possibilities, and together they specify the output uniquely, but without synergy because the constraints are independent.

\textbf{Overjective coverage} creates the potential for synergy, but synergy only manifests when the overlapping ambiguity cannot be independently resolved. If an injective input can disambiguate an overjective input, no synergy occurs---the injective input carries the information independently. Synergy arises when resolving the output requires jointly considering inputs whose supports overlap in ways that cannot be independently untangled.

Looking back at the examples: The pure synergy of XOR results from fully overjective inputs. The mixed contributions of OR reflect a combination of injective and overjective states. The LED square exhibits no synergy despite requiring both inputs because its coverage is fully pseudo-injective where each switch constrains the output to disjoint subsets.

\subsubsection{Coverage Scope}

It is important to distinguish the level, or scope, at which coverage type is assessed. For example, an overjective input may be part of an injective function, depending on how the ambiguity of that input is resolved. We distinguish three scopes: state, input, and functional.

At the state scope, we consider how individual input values map to output values. A state (i.e., a specific value assumed by an input variable) is injective if it leads to a single output, pseudo-injective if it maps to a distinct but multi-valued subset of possible outputs, and overjective if its output subset overlaps with those of other states of the same variable.

At the input scope, we assess how the collection of states belonging to a single input variable partitions the output space. The coverage type of the variable is determined by its most ambiguous state: injective if all states are injective, pseudo-injective if at least one state has multi-valued support but none overlap, and overjective if any state's support overlaps with another.

At the functional scope, we consider the joint mapping of all input variables together. The coverage type of a function is determined by what is required to resolve its outputs: injective if all outputs can be uniquely determined through injective mappings alone, pseudo-injective if some outputs require pseudo-injective mappings but none require resolving overlapping supports, and overjective if any output requires distinguishing between input states whose supports overlap.

\subsubsection{Incorrect model specification leads to misleading decompositions}
The coverage framework also clarifies what happens when a system is misspecified. As discussed in Section~\ref{section:handling-dependent-inputs}, inputs that are not truly independent inflate the entropy of the input space. The coverage perspective reveals why this is problematic: when FID completes the Cartesian product of inputs that are actually dependent, it assigns outputs to combinations that cannot occur in the real system. This can introduce overlapping supports where the true system has none, generating spurious overjectivity---and therefore spurious synergy---in the resulting decomposition. FID faithfully decomposes the system as defined. It is thus critical that dependent inputs are identified and merged as discussed in Section~\ref{section:handling-dependent-inputs}.

\subsubsection{Coverage Type Examples}

\begin{figure}[h]
\centering
\includegraphics[width=0.5\textwidth]{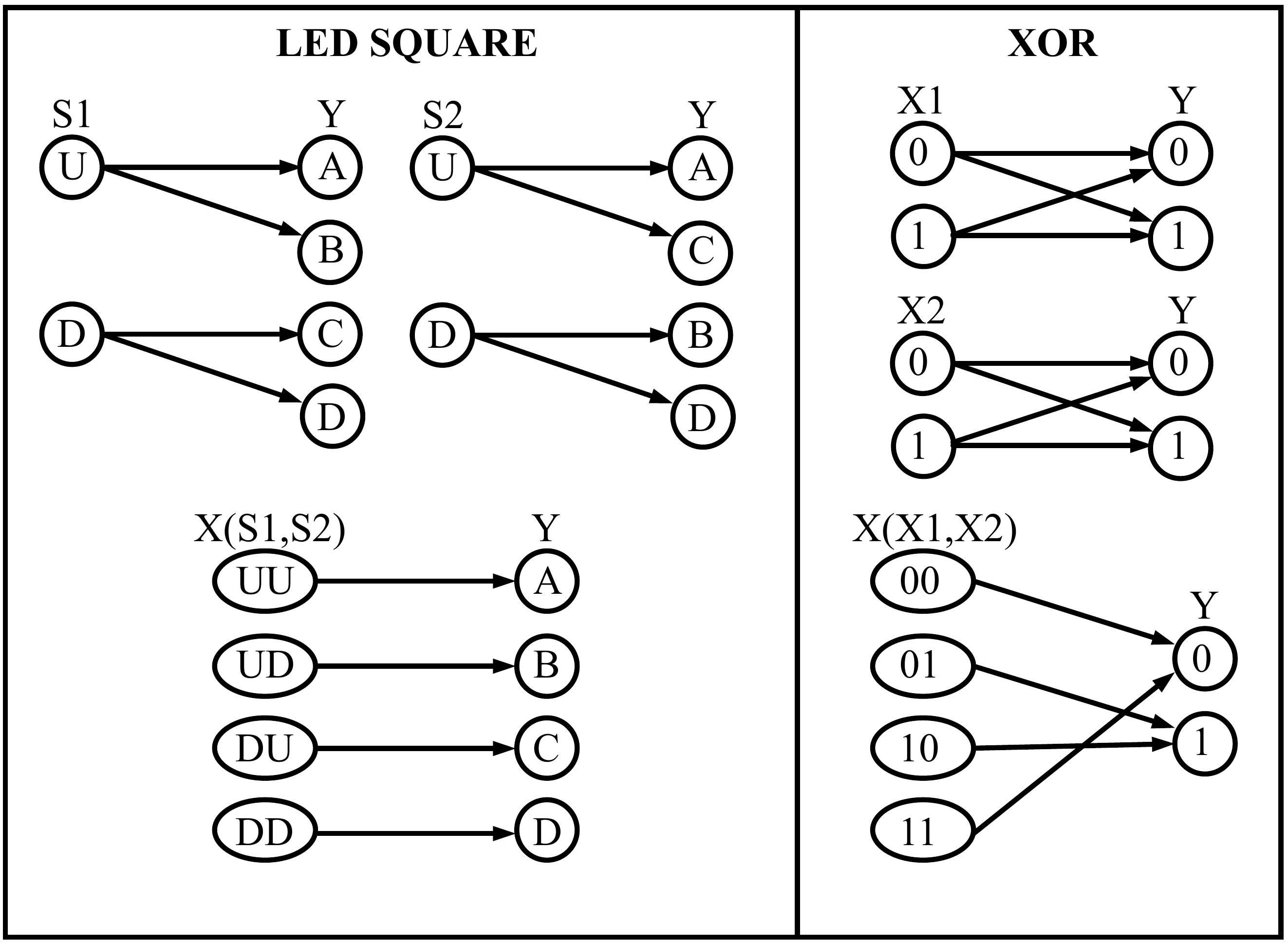}
\caption{In the \textbf{LED Square} example(left), both inputs, S1 and S2, are pseudo-injective, while the function as a whole is injective. In contrast, \textbf{XOR} (right) is overjective in both the input and at the function scopes.}
\label{fig:scope_examples}
\end{figure}

Figure~\ref{fig:scope_examples} illustrates coverage type classification at different scopes for two functions introduced in Section~\ref{full_spec_examples}.

\textbf{LED Square.} At the state scope (top), each switch state maps to a disjoint pair of outputs. Both states are pseudo-injective. At the input scope, each switch variable is therefore pseudo-injective. At the functional scope (bottom), the joint mapping shows that each combination of switch positions uniquely determines a single output: (UU)→A, (UD)→B, (DU)→C, (DD)→D. The function is injective at this scope.

\textbf{XOR.} At the state scope (top), both input states map to overlapping output sets: 0→\{0,1\} and 1→\{0,1\}. All states are overjective. At the input scope, both variables are overjective. At the functional scope (bottom), the joint mapping shows multiple input combinations producing the same output. XOR is thus also overjective at the functional scope.

These two examples illustrate how coverage type can differ across scopes: the LED square is pseudo-injective at the input scope but injective at the functional scope, while XOR remains overjective at every level.

\section{Conclusion}

Functional Information Decomposition defines information decomposition with respect to the system's input-output mapping rather than the observed data distribution. Under complete specification, the decomposition into independent and synergistic components is unique and unambiguous. Under incomplete specification, FID characterizes the space of decompositions consistent with the data, making the limits of inference explicit. The coverage type framework provides a geometric interpretation of these quantities, grounding the distinction between independent and synergistic information in the structure of input-output mappings.

Several directions remain open. The multi-output extension outlined in Section~\ref{sec:multi-output} invites a systematic theory of how decompositions relate across output groupings. The coverage type framework itself may admit further formalization, particularly in connecting geometric properties of input-output mappings to quantitative bounds on FID's information quantities. A Python implementation of FID is available as the open-source package \texttt{fid-tools} (pypi.org/project/fid-tools/).

FID provides a principled foundation for information decomposition within Shannon information theory. Finally, and perhaps most importantly, because FID is conservative, it not only provides sound decompositions but also a clear and honest signal for when the data are insufficient to support meaningful conclusions.

\textbf{Data Availability}: {
The cancer drug response data analyzed in figure~\ref{fig:gene_data} were obtained from the Genomics of Drug Sensitivity in Cancer (GDSC) project, and transcriptomic data were obtained from the DepMap portal. These datasets are publicly available from their respective repositories. No new data were generated in this study.}

\textbf{Code Availability}: {
A Python implementation of Functional Information Decomposition is available as the open-source package fid-tools, which can be installed via pip. The package provides all functionality required to reproduce the analyses presented in this study. Documentation and examples are included with the package.}

\textbf{Acknowledgments}: {The authors would like to thank Jory Schossau and Max Foreback for their insights. Large Language Models (ChatGPT, version 5.2,5.3,5.4; Claude, Sonnet 4.5,4.6 and Opus 4.6; DeepSeek-V3.2) were used to refine, but not to create, content in this work.}

\textbf{Author Contributions}: {C.B. developed the theory, performed the analyses, generated all figures, and wrote the manuscript. V.R., A.H., C.O., E.D. and C.A. contributed to theory development and manuscript editing. C.A. provided data used to generate figure 3. All authors approved the final version.}

\textbf{Conflicts of interest}: {The authors declare no conflicts of interest.} 

\textbf{Research funding}: {This research did not receve finacial support.} 

\bibliographystyle{unsrt}
\bibliography{ref}

\end{document}